# Strong Reduction of Thermal Conductivity of WSe$_2$ with Introduction of Atomic Defects

Bowen Wang[a], Xuefei Yan[a], Hejin Yan[a], Yongqing Cai[a]*

[a]Institute of Applied Physics and Materials Engineering, University of Macau, Macau, China

Abstract

The thermal conductivities of pristine and defective tungsten diselenide (WSe$_2$) are investigated by using equilibrium molecular dynamics method. The thermal conductivity of WSe$_2$ increases dramatically with size below a characteristic with of ~5 nm and levels off for broader samples and reaches a constant value of ~2 W/mK. By introducing atomic vacancies, we discovered that the thermal conductivity of WSe$_2$ is significantly reduced. In particular, the W vacancy has a greater impact on thermal conductivity reduction than Se vacancies: the thermal conductivity of pristine WSe$_2$ reduced by ~60% and ~70% with the adding of ~1% of Se and W vacancies, respectively. The reduction of thermal conductivity is found to be related with the decrease of mean free path (MFP) of phonons in the defective WSe$_2$. The MFP of WSe$_2$ decreases from ~4.2 nm for prefect WSe$_2$ to ~2.2 nm with the adding of 0.9% Se vacancies. More sophisticated types of point defects, such as vacancy clusters and anti-site defects, are explored in addition to single vacancies, and are found to dramatically renormalize the phonons. The reconstruction of the bonds leads to localized phonons in the forbidden gap in the phonon density of states which leads to the drop of thermal conduction. This work demonstrates the influence of different defects on thermal conductivity of single-layer $WSe_2$, providing insight into the process of defect-induced phonon transport as well as ways to improve heat dissipation in $WSe_2$-based electronic devices.

Keywords: molecular dynamics, thermal conductivity, $WSe_2$, atomic defects, phonon,

1. INTRODUCTION

Due to its extraordinary properties[1-3], two-dimensional (2D) transition metal dichalcogenides (TMDs) with sandwich structures have piqued researchers' attention in a variety of fields, including electronics[4-8], optoelectronics[9-11], and thermoelectrics [12-16]. Semiconducting tungsten diselenide ($WSe_2$), one of the TMDs, drawn a lot of interest for thermal management of device due to its ultra-low thermal conductivity[17]. Experimental measurements[18-20] and numerical simulations[21-23] have both applied to explore inherent thermal properties of single-layer $WSe_2$. $WSe_2$ thin films have an in-plane thermal conductivity of around 1.5 W/mK measured by using a suspended device[24]. To comprehensively study the lattice thermal conductivity of $WSe_2$ monolayers, Zhou et al. employed first-principles calculations paired with the phonon Boltzmann transport equation[25], and discovered that the single-layer $WSe_2$ have extremely low thermal conductivity (3.935 W/mK) due to its ultralow Debye frequency and hefty atom mass. Because of its low heat conductivity and direct bandgap semiconducting nature, $WSe_2$ is an attractive candidate for thermoelectric applications.

According to earlier studies[26-30], different types of defects have a considerable impact on mechanical, electrical, and thermal properties. Meanwhile, structural defects were unavoidable and discovered in as-grown 2D materials due to growth and integration[31-33]. In the as-grown single-layer $WSe_2$, for example, structural defects such as point defects are ubiquitous[34, 35]. It is thus critically important to investigate the impact and mechanisms of defects and their impact on $WSe_2$, which is crucial for thermal management in $WSe_2$-based thermoelectric applications. The thermal conductivity of amorphous $WSe_2$ thin films in cross-plane was measured experimentally by Chiritescu et al. to be as low as 0.05 W/mK at ambient temperature due to extensive phonon localization[18]. Disordered $WSe_2$ has a thermal conductivity significantly lower than compacted single-crystal platelets, according to Shi et. al[36]. Previous calculations on the thermal properties of $WSe_2$ were confined to pristine $WSe_2$ configurations[23, 37], despite extensive simulations on defective materials were reported such as graphene[38] and $MoSe_2$[39]. Overall, despite previous research showing that the presence of lattice defects reduces thermal transport characteristics, only a few research have looked into how defects affect thermal transport in single-layer $WSe_2$ and the phonon transport mechanism of defective $WSe_2$ is missing.

In this work, the effects of size and various defects on thermal properties of $WSe_2$ are explored using equilibrium molecular dynamics (EMD) simulations. The mean free path (MFP) of phonons is extrapolated to illustrate the reduction of thermal conductivity explained by the undercoordinated atoms near the vacancies. The formation energies of higher order defects are calculated and their impact on the thermal conductivity of WSe$_2$ are also studied. By examining phonon density of states (PDOS), physical process of thermal conductivity lowering with different defects was investigated.

## 2. METHODOLOGY

The large-scale atomic/molecular massively parallel simulator (LAMMPS)[40] was used to run all of our molecular dynamics (MD) simulations in this investigation. We employ the Stillinger-Weber (SW) reported by Norouzzadeh et. al[23] to describe interatomic interactions, which reproduces the phonon dispersion in close to first-principles calculations[41] and obtain in-plane thermal conductivity results fitting well with experiments[25]. The formation energies of point defects seen in single-layer WSe$_2$ that we obtained from this potential accord well with DFT simulations on a quantitative level[42]. The Fig. 1 illustrates the hexagonal sandwiched form of the WSe$_2$ structure from the top and side. In the x (zigzag) and y (armchair) directions, the periodic boundary condition was utilized, whereas the free boundary condition was used in the z direction. Each structure is initially equilibrated for 100000 time steps at constant volume and 300 K using a time step of 0.5 fs. After the equilibration, we proceed with a constant energy run NVE for up to $1 \times 10^8$ time steps. To make a more consistent analysis, for each content of defect, five independent configurations with different length and randomly produced defective configurations are calculated and the values are averaged. It was reported[43] that that in single-layer WSe$_2$, in-plane thermal conductivity in the armchair direction is kind of stronger than in the zigzag direction, and this anisotropy being more prominent at low temperatures. In this work, by using EMD method, we mainly focus on the thermal conductivity along the armchair directions at 300 K. The principles of fluctuation dissipation and linear response[44] are utilized to compute thermal conductivity within the EMD paradigm[45]. The heat flux vectors, as well as their correlations, are computed during the simulation. The thermal conductivity in the armchair direction $\kappa_{yy}$, is connected to the heat current autocorrelation function (HCACF) by Green-Kubo expression[44, 46]

$$\kappa_{yy} = \frac{1}{V k_B T^2} \int_0^\tau \langle j_y(t) \cdot j_y(0) dt \rangle, \qquad (1)$$

where the $V$, $k_B$, T represent system volume, Boltzmann constant, system temperature respectively. The angular brackets indicate an ensemble average of the heat flux $j$ autocorrelation.

We calculate the PDOS of WSe$_2$ with different vacancy configurations by taking the Fourier transform of the velocity autocorrelation function in order to depict the phonon vibrations at different frequencies:

$$F(\omega) = \frac{1}{\sqrt{2\pi}} \int_{-\infty}^{\infty} e^{-i\omega t} \langle v(t) \cdot v(0) \rangle / (\langle v(0) \rangle \cdot v(0)) dt, \qquad (2)$$

where $F(\omega)$ represents the PDOS at $\omega$ angular frequency, $v(t)$ and $v(0)$ represent the atomic velocity vectors at the time of $t$ and 0, respectively.

## 3. RESULTS AND DISCUSSION

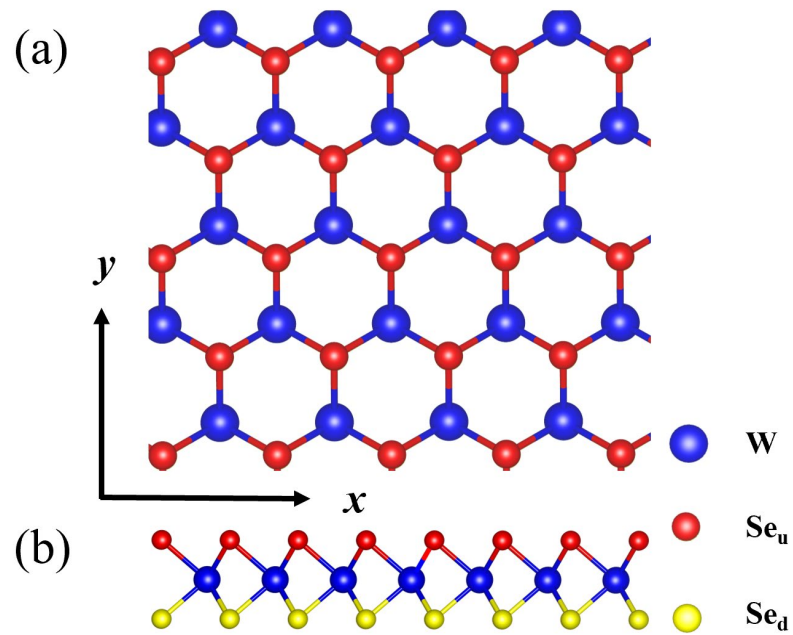

**Figure 1.** Top and side views of single-layer WSe$_2$ configuration. The top and bottom layers' selenide atoms are shown by red and yellow balls, respectively and the blue balls represent tungsten atoms.

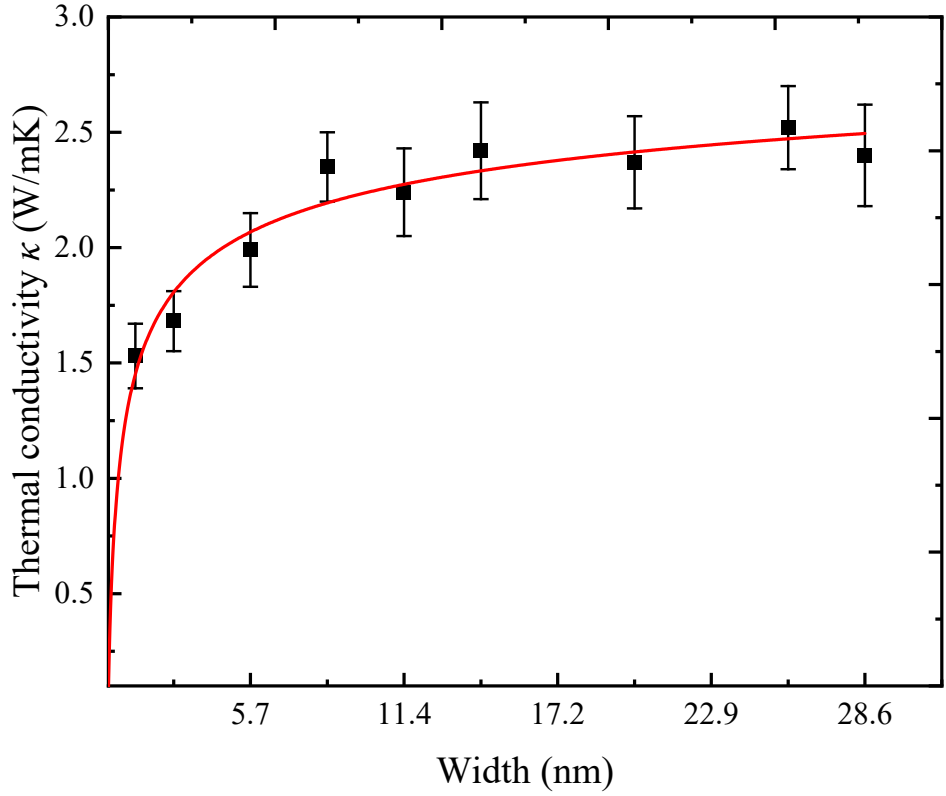

**Figure 2.** Thermal conductivity as a function of width for thermal conductivity in the armchair direction with a length of 50 nm. The point corresponds to the data points while the solid line represents the fitted curve.

Firstly, the effect of width (normal to the transport direction) of the model on thermal conductivity is examined. The length of all simulated single-layer $WSe_2$ was fixed to 50 nm which is wide enough to resolve the Brillion zone while also avoiding a thermal conductivity forecast divergence. Fig.2 depicts the width dependence of thermal conductivity for single-layer $WSe_2$ with considering width ranging from 3 to 28.6 nm. To suppress the uncertainty arisen from the statistical thermal fluctuations which are likely to occur in the calculation of thermal conductivity using molecular dynamics[47], 10 independent simulations for various cell widths are performed. We can observe that when the width rises, the thermal conductivity increases as well. The thermal conductivity of WSe$_2$ increases dramatically with size below a characteristic with of ~5 nm and levels off for broader samples and reaches a constant value of ~2 W/mK. The thermal transport here is governed by Umklapp scattering and boundary scattering, and the latter is more dominant than Umklapp scattering when the width is narrow. The thermal conductivity of WSe$_2$ increases as the width grows due to a reduction in scattering from the edge localized phonon effect. The Umklapp scattering becomes the leading factor that governs the phonon scattering while the boundary scattering becomes gradually weakened.

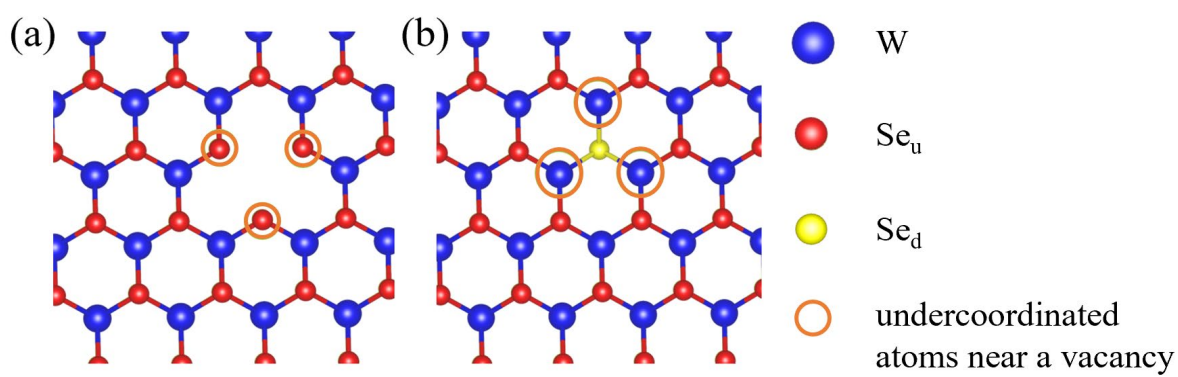

**Figure 3.** Atomic configurations for single tungsten vacancy (a) and selenium vacancy (b) in single-layer WSe$_2$.

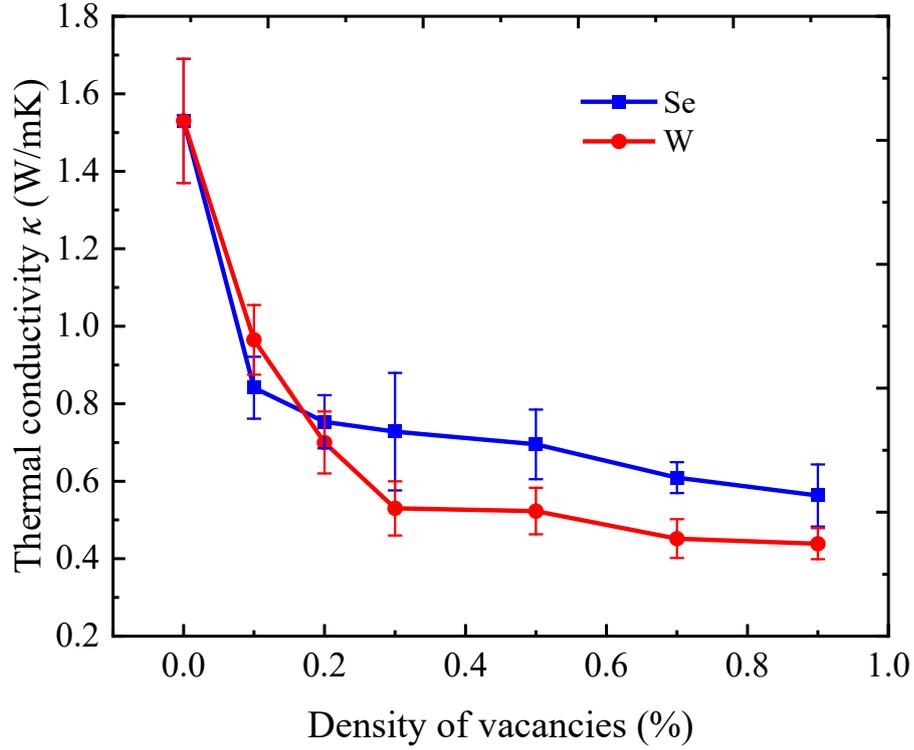

**Figure 4.** Thermal conductivity of single-layer WSe$_2$ in relation to defect ratio of W and Se vacancies.

To see how defects affect phonon transport in WSe$_2$ monolayers, the in-plane size of the simulation box is fixed as 50 × 10 nm and five independent simulations with randomly produced vacancies are carried for a specific content of vacancy. Both of the thermal conductivity of pristine single-layer WSe$_2$ (1.53 ± 0.16 W/mK) calculated by EMD method and MFP extrapolated (4.18 ± 0.08 nm) fits well with NEMD results reported by Norouzzadeh[23]. Trends in thermal conductivity with two different vacancies: Se vacancy (V$_{Se}$) and W vacancy (V$_W$), are shown in Fig. 4. The thermal conductivity of pristine WSe$_2$ fell sharply as the defect concentration rises from 0% to

0.9% and reduced by ~60% and ~70% when the density of vacancies equal to 0.9% for $V_{Se}$ and $V_W$, respectively. The significant influence of defects on thermal conductivity is also found in $MoS_2$[48] and $MoSe_2$[49]. The $V_W$ has a greater impact on the thermal conductivity of $WSe_2$., similar to the case of $MoS_2$ with lattice defects [50] . Reduction of relaxation time (τ) and mean free path (MFP), which caused by missing atoms, linkages, and different force constants near vacancies, are the main causes of thermal conductivity reduction.

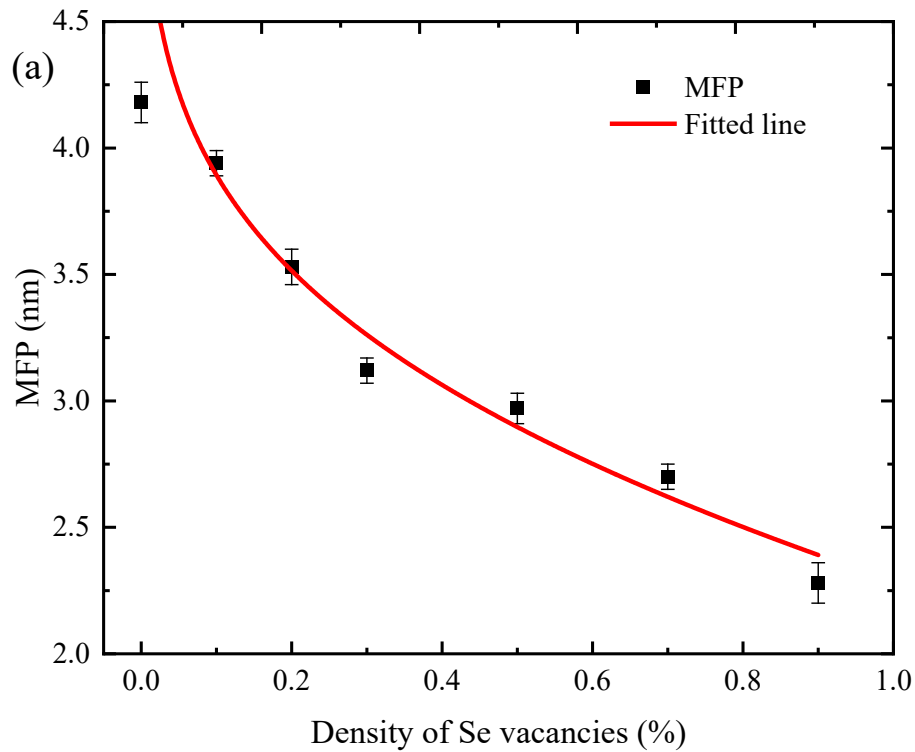

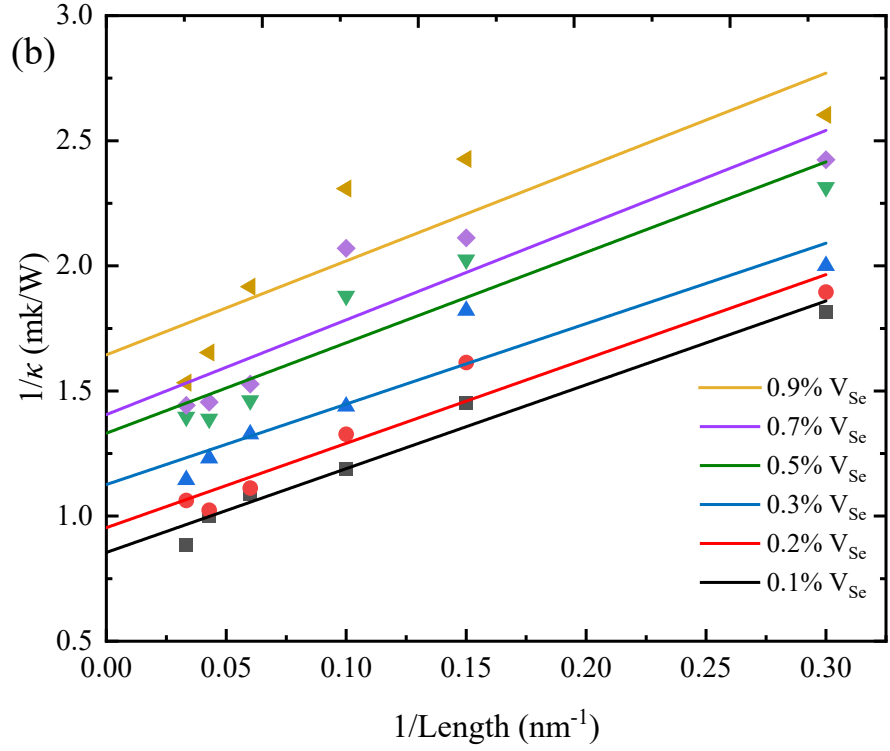

**Figure 5.** (a) Mean free path of WSe$_2$ in relation to density of V$_{Se}$ at 300 K. (b) The inverse of thermal conductivity versus inverse of length in WSe$_2$ containing different contents of V$_{Se}$.

The following formula relates the thermal conductivity of a single layer material with a limited length to the thermal conductivity of an endlessly long single-layer [51]:

$$\frac{1}{\kappa} = \frac{1}{\kappa_\infty}(\frac{1}{L} + 1) \qquad (3)$$

Where $L$ is the phono MFP. The thermal conductivity of finite length and infinitely long single-layer are represented by $\kappa$, $\kappa_\infty$, respectively. As a result, single-layer thermal conductivity's inverse is proportional to the inverse of their lengths. The MFP of defective single-layer WSe$_2$ is extrapolated in Fig. 5(a) and Fig. 5(b) illustrates the inverse of thermal conductivity versus inverse of single-layer lengths. Here we only examine V$_{Se}$ which is more likely to be formed owing to a lower formation energy than V$_W$ as we will show below. The MFP of WSe$_2$ is found to decrease from ~4.2 nm for prefect WSe$_2$ to ~2.2 nm with the adding of Se vacancy at 0.9% V$_{Se}$. Because the MFP of a phonon is inversely related to its scattering rate, a smaller MFP indicates an increase in scattering rate and a drop in heat conductivity.

Matthiessen's rule[52] states that the total scattering rate $\tau^{-1}$ is written as:

$$\tau^{-1} = \tau_{Anh}^{-1} + \tau_B^{-1} + \tau_V^{-1} + \tau_F^{-1} \qquad (4)$$

where $\tau_{Anh}^{-1}$ and $\tau_B^{-1}$ represent the Umklapp phonon-phonon and phonon-boundary scattering rate respectively, $\tau_V^{-1}$ is related to loss of atom near the vacancy, and $\tau_F^{-1}$ is induced by the shift in force constants between the undercoordinated atoms near the vacancies. $\tau_V^{-1}$ can be written as[53, 54]:

$$\tau_V^{-1} = x(-\frac{M_V}{M} - 2)^2 \frac{\pi}{2} \frac{g(\omega)\omega^2}{N} \qquad (5)$$

where $x$ is the defect concentration, $M$ is the atomic mass average, $M_V$ is the absent atom's mass at the void, $g(\omega)$ is the phonon DOS and $N$ is atoms number.

According Eq. 5, $\tau_V^{-1}$ induced by V$_W$ is greater than that by V$_{Se}$ since the atomic masses of the W (183.84) is significantly high than Se (78.971) atoms. This is true in the sense in Fig. 4 that V$_W$ affects thermal conductivity more than V$_{Se}$ as the variation in $g(\omega)$ and $G$ caused by vacancies are small because of the small concentrations of vacancies in our study.

$\tau_F^{-1}$ can be written as:

$$\tau_F^{-1} = nx(\frac{\delta k}{k})^2 4\pi \frac{\omega^2 g(\omega)}{G} \qquad (6)$$

where $n$ is the number of atoms located around the vacancy, $x$ and $\delta k$ are force constants before and after the change. For single $V_{Se}$ and single $V_W$, n equals 3 and 6, respectively, as illustrated in Fig. 3. The $\delta k$ caused by the $V_W$ is bigger than that caused by the $V_{Se}$ due to higher atomic mass of W. According to Eq. 6, the scattering rate induced by variation of force constants around vacancies $\tau_F^{-1}$ is greater induced by the $V_W$.

In addition to the $V_{Se}$ and $V_W$ single-vacancy-type defects, we also calculate other types of vacancies which are possible in the WSe$_2$. Zhang et. al[34] found that the W related defects are likely to occur and three possible defect structures are proposed including the anti-site of W and Se atom (Se$_W$) in which a Se atom replaces a W atom, vacancy complex of three missing Se atoms near the W vacancy site ($V_{WSe3}$) and vacancy complex of three missing Se atoms near the W atom ($V_{Se3}$). Besides, other types of vacancies: the two Se vacancies in an apical form ($V_{Se2}$) and the six vacancies arranged in chain ($V_{Se6C}$) and triangle ($V_{Se6T}$) forms. The latter configurations were observed in WS$_2$ by using scanning transmission electron microscopy[31]. The vacancy structures are shown in Fig. 6.

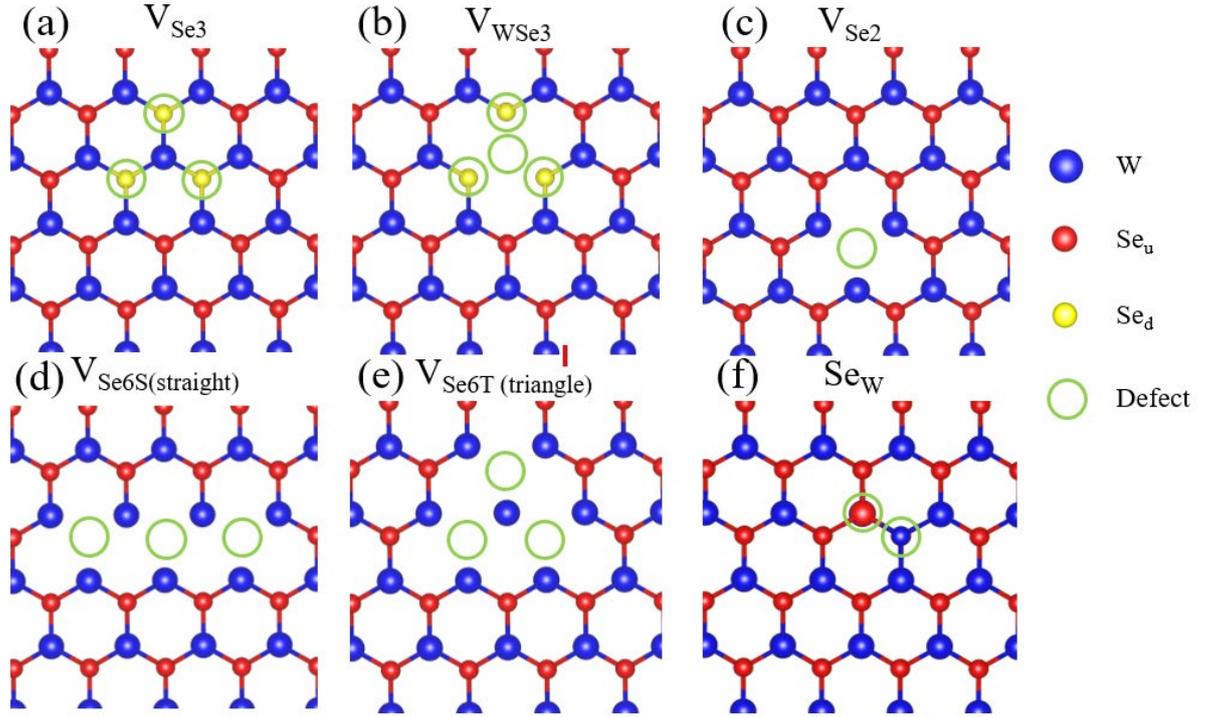

**Figure 6.** Six kinds of vacancies configuration in single-layer WSe$_2$: (a) Vacancy complex of three Se atoms, (b) Vacancies of single W and nearby three Se vacancies, (c) Vacancy of double Se atoms (in the same column), (d) Vacancy complex of six missing Se atoms (triangle), (e) Vacancy complex of six missing Se atoms (chain) (f) Anti-site defect of W and Se atom.

We performed MD calculations for the formation energy of various vacancies to compare the stability. The formation energy $E_v^f$ is calculated from the following expression:

$$E_v^f = E_f - \left[\frac{N_0 - N_f}{N_0}\right] * E_i \qquad (7)$$

where $E_f$ and $E_i$ represent the final energy after the atoms is removed from the cell, the initial energy of the perfect system, respectively, and $N_0$ and $N_f$ are the total number of atoms in the cell before and after the atoms being removed.

Table 1. The formation energy of different types of defects in WSe$_2$.

| Defects type | $E_v^f$(this work) | $E_v^f$(DFT)[55] | $E_v^f$(ReaxFF)[55] |
|---|---|---|---|
| V$_W$ | 5.71 | 5.26 | 5.37 |
| V$_{Se}$ | 2.27 | 2.66 | 2.34 |
| V$_{Se2}$(separate) | 4.55 | - | - |
| V$_{Se2}$(in the same column) | 4.17 | 4.82 | 5.17 |
| V$_{Se3}$ | 6.51 | - | - |
| V$_{Se6C}$ | 12.95 | - | - |
| V$_{Se6T}$ | 13.87 | - | - |
| V$_{WSe3}$ | 12.56 | 10.92 | 11.71 |
| Se$_W$ | 5.49 | 4.51 | 4.04 |

The formation energy of various vacancies investigated in this work are calculated in Table 1. Meanwhile, the formation energies of some type defects calculated by DFT and ReaxFF potential[55] are also listed as comparison. As we can see from the Table 1, a single V$_{Se}$ has the lowest formation energy and the similar results also found in other TMD materials[56], which is also consistent with DFT calculations[34]. Three W atoms relax towards the vacancy site when one Se atom is missing while six dangling Se atom connected to W, which impact the relaxation of these Se atoms. The V$_{Se2}$ defect containing two Se atoms arranged in an apical form has a much lower formation energy than two independent V$_{Se}$, implying a high probability of aggregation of vacancies.

Chain vacancies are favored over isolated vacancies in higher order vacancies, which is observed by using transmission electron[31]. Comparison of two possiblie configurations for $V_{Se6}$: vacancy complex of six Se atoms in triangle form $V_{Se6T}$ and linear chain $V_{Se6C}$, shows that the latter is more stable by 0.92 eV.

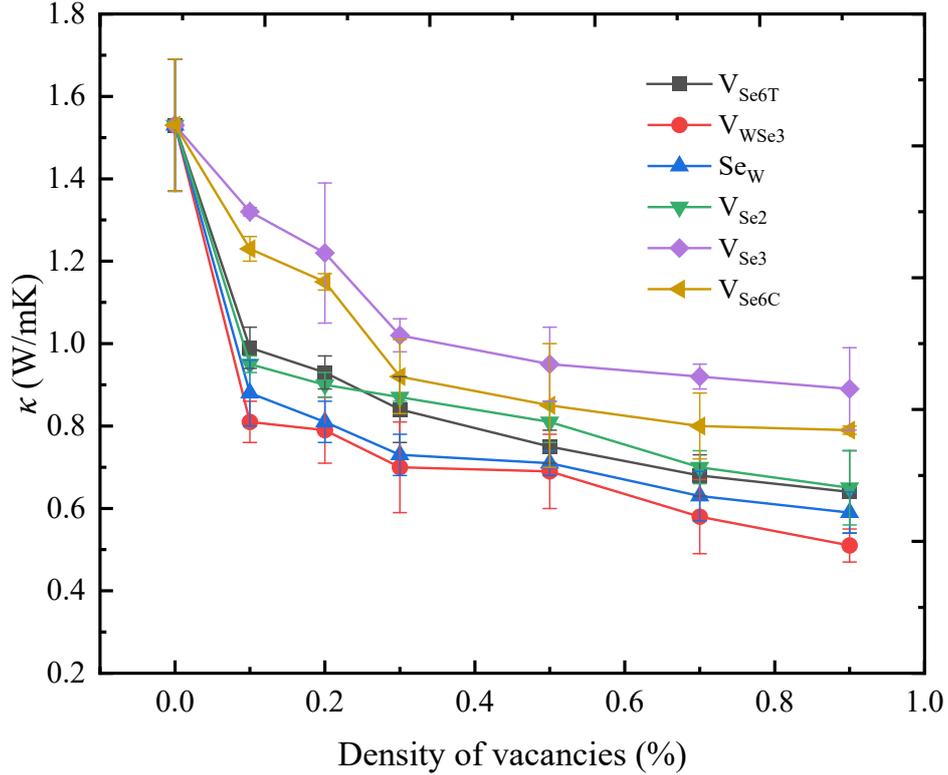

**Figure 7.** Thermal conductivity of single-layer $WSe_2$ as a function of the density of vacancies with several types of vacancies ($Se_W$, $V_{WSe3}$, $V_{Se3}$, $V_{Se2}$, $V_{Se6C}$, $V_{Se6T}$).

To investigate the effect of content of each type of defect on single-layer $WSe_2$, five independent simulations with randomly produced defective configurations for a specific content are calculated and in a simulation box fixed as 50 × 10 nm. The averaged thermal conductivity is obtained to ensure a coherent result and reduce the divergency. As seen in Fig. 7, the thermal conductivity of single-layer $WSe_2$ falls as vacancies are added and the reduction of the thermal conductivity is influenced by the vacancies type. Because of the increased number of undercoordinated atoms around a vacancy generated by trigonal symmetry (7 and 9 for $V_{Se6C}$ and $V_{Se6T}$, respectively), the loss of thermal conductivity is greater with $V_{Se6T}$ than $V_{Se6C}$ in the chain vacancies we examined. The anti-site defect $Se_W$ leads to a higher reduction of thermal conductivity due to a strong renormalization of phonons which will be analysised by PDOS later. It is also found in first-principles calculation[34] that the trigonal symmetry in $WSe_2$ with $Se_W$ defect is broken during structural relaxation process which may be the reason why the thermal conductivity of $WSe_2$ is highly suppressed to the presence of $Se_W$ defects.

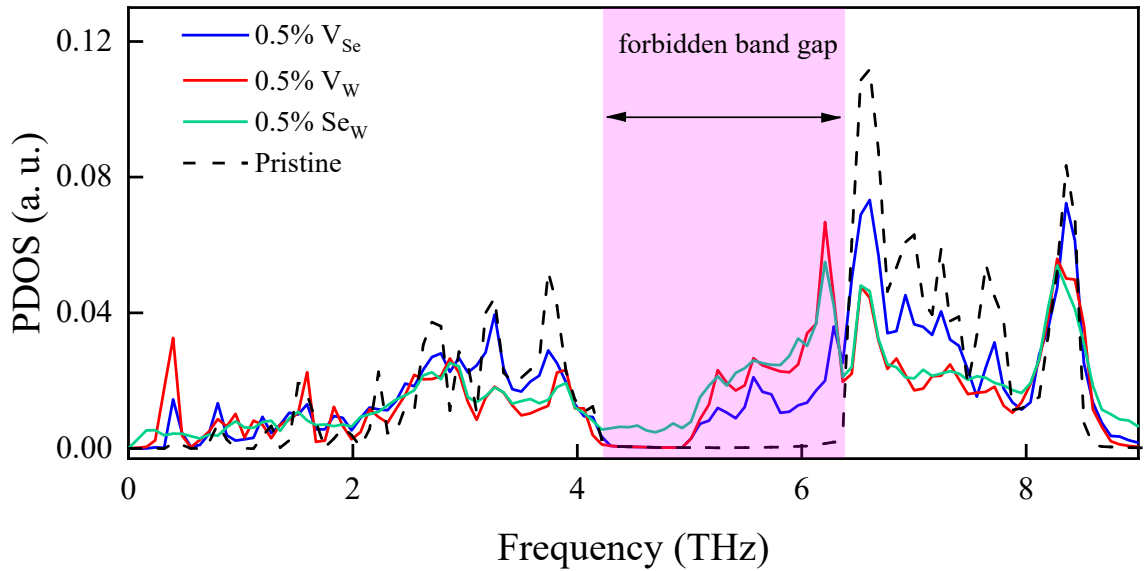

**Figure 8.** Phonon DOS of single-layer pristine and defective $WSe_2$ (0.5% $V_W$, $V_{Se}$, and $Se_W$)

We calculate the PDOS of single-layer $WSe_2$ with 0.5% single $V_W$, $V_{Se}$, and $Se_W$ defects respectively to illustrate how the vacancy influences the phonon scattering of

WSe$_2$. The PDOS would be particularly useful for revealing the alteration of the spectrum of vibrations upon the introduction of bond breaking and formation. In our calculation, a same size of the supercell of WSe$_2$ is used to allow a direct comparison. We find that all the defects are associated with broadened phonon peaks, especially for those optical branches, while simultaneously reduce the peak intensities in the PDOS. This indicates the bond breaking and reorganization induced by the depletion or replacement of atoms in the defective core. Moreover, the forbidden band gaps between the acoustic and optical branches become narrow in the presence of a defect, compared to that for the pristine WSe$_2$. Particularly, the gap significantly vanishes in the cases of 0.5% anti-site Se$_W$ defect. These changes reflect the strong potential gradient around the defective core which further explains a decrease in the lifetime of the phonon[57], and accordingly the drop of the thermal conductivity of the WSe$_2$.

## 4. SUMMARY

In conclusion, we investigated the effects of width and various single and higher order vacancies on the thermal conductivity of single-layer WSe$_2$ using EMD method. The thermal conductivity of WSe$_2$ increases monotonically with size below a critical size of ~5 nm after which the thermal conductivity increases slowly and gradually reaches a constant. It is found that the thermal conductivity of single-layer WSe$_2$ is significant suppressed by the addition of vacancies defects. The MFP of defective WSe$_2$ is extrapolated and the decline of MFP is responsible for the reduction of thermal conductivity. Moreover, the decreased thermal conductivity also depends on vacancy type due to the different missing atom mass and undercoordinated atoms near the vacancies. Higher order defects and the effect on the thermal conductivity are investigated. We found that the impact of the vacancies depends on the missing atom mass, undercoordinated atoms near the vacancies, and the stability of the defective structure. Moreover, the analysis of PDOS of defective $WSe_2$ reveals that decline in the value of PDOS peaks and shrinkage of forbidden band gaps are accompanied with the reduction of thermal conductivity.

## ACKNOWLEDGMENT

This work was supported by the University of Macau (SRG2019-00179-IAPME) and the Science and Technology Development Fund from Macau SAR(FDCT-0163/2019/A3), the Natural Science Foundation of China (Grant 22022309) and Natural Science Foundation of Guangdong Province, China (2021A1515010024). This work was performed in part at the High-Performance Computing Cluster (HPCC) which is supported by Information and Communication Technology Office (ICTO) of the University of Macau.

## DATA AVAILABILITY

The data that support the findings of this study are available from the corresponding author upon reasonable request.

## CONFLICT OF INTEREST

The authors have no conflicts to disclose.